\documentclass[a4paper]{llncs}
\usepackage{algorithm2e}
\usepackage{float}
\usepackage{amssymb,amsmath}
\usepackage{graphicx}
\usepackage{subfig}
\usepackage[T1]{fontenc}
\usepackage{amsmath}

\usepackage{amssymb}
\usepackage[utf8]{inputenc}
\usepackage{url,bm}
\urldef{\mailsa}\path|{alfred.hofmann, ursula.barth, ingrid.haas, frank.holzwarth,|
\urldef{\mailsb}\path|anna.kramer, leonie.kunz, christine.reiss, nicole.sator,|
\urldef{\mailsc}\path|erika.siebert-cole, peter.strasser, lncs}@springer.com|    
\newcommand{\keywords}[1]{\par\addvspace\baselineskip
\noindent\keywordname\enspace\ignorespaces#1}

\begin{document}

\mainmatter  

\title{Prediction of severity and treatment outcome for ASD from fMRI} 
\titlerunning{Prediction of severity and treatment outcome for ASD from fMRI}

%
%
%
%
\author{ Juntang Zhuang$^1$, \quad Nicha C. Dvornek $^{2,3}$,\quad  Xiaoxiao Li$^1$, \quad Pamela Ventola$^2$,\qquad James S. Duncan$^{1,3,4}$}
\footnotesize{
 \institute{$^{1}$ Biomedical Engineering, Yale University, New Haven, CT USA  \\ $^{2}$ Child Study Center, Yale University, New Haven, CT USA\\ $^{3}$ Radiology \& Biomedical Imaging, Yale School of Medicine, New Haven, CT USA \\  $^{4}$ Electrical Engineering, Yale University, New Haven, CT USA}
}


%
%
\maketitle
\begin{abstract}
Autism spectrum disorder (ASD) is a complex  neurodevelopmental syndrome. Early diagnosis and precise treatment are essential for ASD patients. Although researchers have built many analytical models, there has been limited progress in accurate predictive models for early diagnosis. In this project, we aim to build an accurate model to predict treatment outcome and ASD severity from early stage functional magnetic resonance imaging (fMRI) scans.    
The difficulty in building large databases of patients who have received specific treatments and the high dimensionality of medical image analysis problems are challenges in this work.
We propose a generic and accurate two-level approach for high-dimensional
regression problems in medical image analysis. First, we perform region-level feature selection using a predefined brain parcellation. Based on the assumption that voxels within one region in the brain have similar values, for each region we use the bootstrapped mean of voxels within it as a feature. In this way, the dimension of data is reduced from number of voxels to number of regions. Then we detect predictive regions by various feature selection methods. Second, we extract voxels within selected regions, and perform voxel-level feature selection. 
To use this model in both linear and non-linear cases with limited training examples, we apply two-level elastic net regression and random forest (RF) models respectively. 
To validate accuracy and robustness of this approach, we perform experiments on both task-fMRI and resting state fMRI datasets. Furthermore, we visualize the influence of each region, and show that the results match well with other findings.
\end{abstract}
\section{Introduction}
\label{sec:intro}
\par Autism spectrum disorder (ASD) is a neurodevelopmental syndrome characterized by impaired social interaction, difficulty in communication and repetitive behavior. ASD is most commonly diagnosed with a behavioral test \cite{baird2003diagnosis}, however, the behavioral test is insufficient to understand the mechanism of ASD. Functional magnetic resonance imaging (fMRI) has been widely used in research on brain diseases and has the potential to reveal brain malfunctions in ASD.
\par Behavior based treatment is a widely used therapy for ASD, and Pivotal Response Treatment (PRT) is empirically-supported \cite{koegel1999pivotal}. PRT addresses core deficits in social motivation to improve social communication skills. Such therapies require large time commitments and lifestyle changes. However, an individual's response to PRT and other behavioral treatments vary, yet treatment is mainly assigned by trial and error. Therefore, prediction of treatment outcome during early stages is essential.
\par fMRI measures blood oxygenation level dependent (BOLD) signal and reflects brain activity. Recent studies have applied fMRI in classification of ASD and identifying biomarkers for ASD \cite{anderson2011functional}. 
Although some regions are found to have higher linear correlations with certain types of ASD severity scores, the correlation coefficient is typically low (below 0.5). Moreover, most prior studies apply \textit{analytical} models, and lack \textit{predictive} accuracy.
\par
The goal of our work is to build accurate  \textit{predictive} models for fMRI images. To deal with the high dimensionality of the medical image regression problem, we propose a two-level modeling approach: 1) region-level feature selection, and 2) voxel-level feature selection. In this paper, we demonstrate predictive models for PRT treatment outcomes and ASD severity, and validate robustness of this approach in both task fMRI and resting state fMRI datasets. Furthermore, we analyze feature importance and identify potential biomarkers for ASD.   
\section{Methods}
\subsection{Two-level modeling approach}
\label{pipe}
\par
\begin{figure}[t]%
    \centering
    \includegraphics[width=\linewidth]{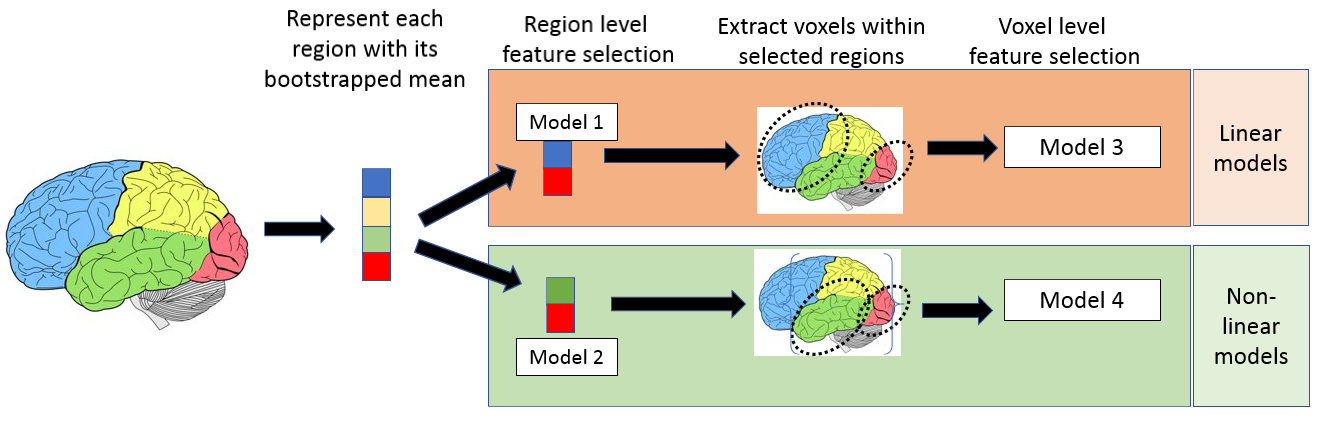} 
    \caption{
    {\small{
    Flowchart of the proposed approach. Each column represents a stage of the approach, region-level and voxel-level models. Top row shows linear models (e.g. elastic net regression), bottom row shows non-linear models (e.g. random forest).}}%
    }
    \label{flow}
\end{figure}

 Dimensionality of medical images (i.e., the number of voxels) is far higher than the number of subjects in most  medical studies. The high dimensionality causes inaccuracy in variable selection and affects modeling performance. However, medical images are typically locally smooth, and voxels are not independent of each other. This enables us to perform the following two-level feature selection as shown in  Fig. \ref{flow}. The proposed procedure first selects important features at the region level, then performs feature selection at the voxel level. Our generic approach can be used with both linear and non-linear models.
\vspace{-0.4cm}
\subsubsection{Region-level modeling and variable selection}
\hfill\\
Based on brain atlas research, we assume that voxels within the same region of a brain parcellation have similar values. Therefore, we use the bootstrapped (sample with replacement) mean for each region as a feature, reducing dimension of data from number of voxels to number of regions. Then we can perform feature selection on this new dataset, where each predictor variable represents a region.
\par
Beyond dimension reduction, representing each region with the bootstrapped mean of its voxel values decreases correlation between predictor variables. Another potential benefit is to increase sample size. We can generate many artificial training examples from one real training example by repeatedly bootstrapping each region. Since this generates correlated training samples, repeated bootstrapping can only be used in models that are robust to sample correlation.
\vspace{-0.5cm}
\subsubsection{Voxel-level modeling and variable selection}
\hfill\\
Region-level feature selection preserves predictive regions. However, representing all voxels within a region as one number is too coarse, and may affect model accuracy. Therefore, we extract all voxels within the selected regions, perform spatial down-sampling by a factor of 4, and apply feature selection on voxels.
\vspace{-0.4cm}
\subsubsection{Pipeline repetition}
\hfill\\
Due to the randomness in bootstrapping for region level modeling, we repeat the whole process. For each of the four models (linear and non-linear models at region-level and voxel-level respectively), we average outcomes to generate stable predictions. 
\subsection{Linear and non-linear models}
We can apply any model in the approach proposed in section \ref{pipe}. To instantiate a generic approach for both linear and non-linear cases, we train elastic net regression and random forest (RF) independently, both trained at two levels. 
\vspace{-0.4cm}
\subsubsection{Variable selection with elastic net regression}
\hfill\\
Elastic net is a linear model with both $l1$ and $l2$ penalty to perform variable selection and shrinkage regularization \cite{zou2005regularization}. Given predictor variables $X$ and targets $y$, the model is formalized as
\begin{equation}
\hat{\beta} = \operatorname{argmin}_{\beta} \big \{ \vert \vert y - X\beta \vert\vert ^2 + \lambda \big [ \alpha \vert\vert \beta\vert\vert_1 + (1-\alpha) \vert \vert\beta \vert\vert^2 \big ] \big \}
\end{equation}
where $0\leq \alpha \leq 1$, $\alpha$ controls the proportion of regularization on $l1$ and $l2$ term of estimated coefficients, and $\lambda$ controls the amplitude of regularization. $l2$ penalty is shrinkage regularization and improves robustness of the model. $l1$ penalty controls sparsity of the model. By choosing proper parameters, irrelevant variables will have coefficients equal to 0, enabling variable selection.
\vspace{-0.4cm}
\subsubsection{Variable selection with random forest} 
\hfill\\   
Random forest is a powerful model for both regression and classification problems and can deal with interaction between variables and high dimensionality \cite{liaw2002classification}. Although random forest can handle medium-high dimensional problems, it's insufficient to handle ultra-high dimensional medical image problems. Therefore, two-level variable selection is still essential.
\par
\begin{figure}[!htb]%
    \centering
    \includegraphics[width=12.5cm]{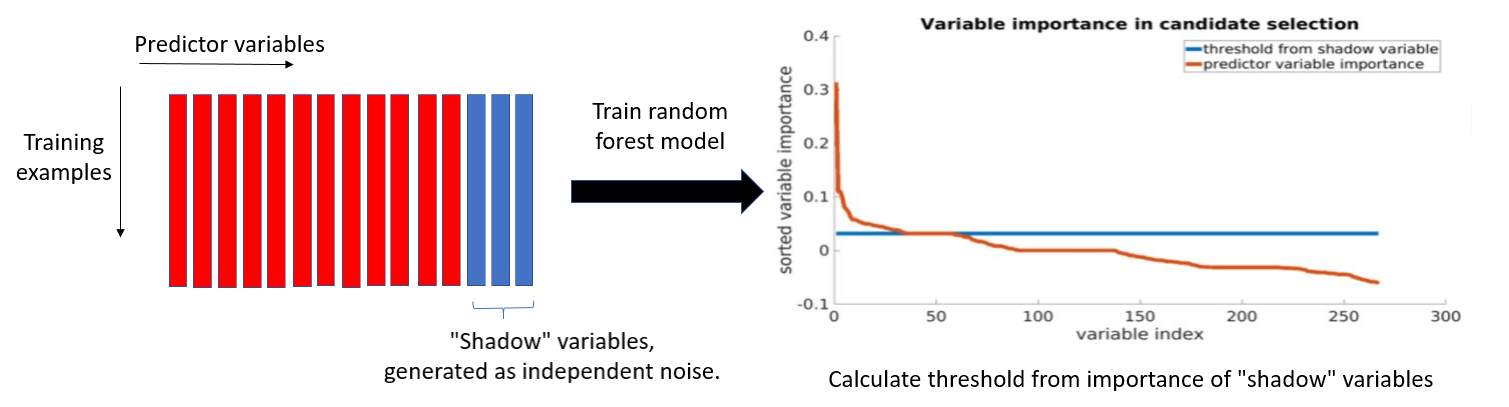} 
    \caption{
    {\small{
    Flowchart of variable selection with random forest and "shadow" method.}}%
    }
    \label{shadow}
\end{figure}

Conventional variable selection technique for random forest builds a predictive model with forward stepwise feature selection \cite{genuer2010variable}. For high dimensional problems, it is computationally intensive. Therefore, we use a similar thresholding method to perform fast variable selection as in \cite{zhuang2018prediction} (Fig. \ref{shadow}). We generate noise ("shadow") variables from a Gaussian distribution independent of target variables. Shadow variables are added to the original data matrix, and a random forest is trained on the new data matrix. The random forest model calculates the importance of each variable. A predictive variable should have higher importance than noise variables. A threshold is calculated as:
\begin{equation}
Thres = \operatorname{median} (VI_i^{shadow})\ \ s.t.\ \  VI_i^{shadow}>0, \ \ i=1,2,...n
\end{equation}
where $n$ is the total number of shadow variables, and $VI_i^{shadow}$ is the importance measure for the $i$th shadow variable.  We use permutation accuracy importance measure \cite{genuer2010variable} in this experiment. The threshold is calculated using positive shadow variable importance because permutation accuracy importance can be negative. We use the median to make a conservative threshold, because even noise variables can have high importance in high-dimensional problems, due to randomness of the model. After variable selection, we build a gradient boosted regression tree model based on selected variables.
\vspace{-0.5cm}
\vspace{-1cm}
\subsection{Visualization of each variable's influence}
To achieve both predictability and interpretability, we use the following methods to visualize influence of each region in the brain. For linear models, we plot the linear coefficients map. 
\par
For non-linear models, we visualize the influence based on the partial dependence plot. The partial dependence plot shows the dependence between target and predictor variables, marginalizing over all other features \cite{friedman2001greedy},  
\begin{equation}
D_l (z_l) = \mathbb{E}_x (\hat{F}(x) \vert z_l) = \int \hat{F}(x) p(z_{\textbackslash l} \vert z_l) dz_{\textbackslash l}
\end{equation}
where $D_l(z_l)$ is the partial dependence function for variable $z_l$, $\hat{F}(x)$ is the trained model, $z_{\textbackslash l}$ is the set of variables except $z_l$, $ p(z_{\textbackslash l} \vert z_l) $ is the distribution of $z_{\textbackslash l}$ given $z_l$. Each $D_l(z_l)$ is calculated as a sequence varying with $z_l$ in practice. 
\par
The influence of each variable is stored in a sequence. For visualization, we summarize the influence of a variable ($\operatorname{Influence}_l$) by calculating the variance of  $D_l(z_l)$ to measure the amplitude of its influence, and the sign of its correlation with $z_l$ to show if it has a positive or negative influence on targets:
\begin{equation} \label{influence}
\operatorname{Influence}_l = \operatorname{Sign} \Big(\operatorname{corr}  \big(D_l(z_l), z_l\big)\Big) \operatorname{Var} \big (D_l(z_l) \big).
\end{equation}
\vspace{-1cm}
\section{Experiments and results}
\subsection{Task-fMRI experiment}
\label{task_experiment}
Ninteen children with ASD participated in 16 weeks of PRT treatment, with pre-treatment and post-treatment social responsiveness scale (SRS) scores \cite{bruni2014test}, and pre-treatment  autism diagnostic observation schedule (ADOS) \cite{lord2000autism} scores measured. Each child underwent a pre-treatment baseline task fMRI scan (BOLD, TR = 2000ms, TE = 25ms, flip angle =  $60^{\circ}$, 
slice thickness = 4.00mm, voxel size $3.44 \times 3.44 \times 4mm^3$) and a structural MRI scan (T1-weighted MPRAGE sequence, TR = 1900ms, TE=2.96ms, flip angle = $9^{\circ}$, slice thickness = 1.00mm, voxel size = $1\times 1 \times 1mm^3$) on a Siemens MAGNETOM Trio TIM 3T scanner. 
\par 
During the fMRI scan, coherent
(BIO) and scrambled (SCRAM) point-light biological motion
movies were presented to participants in alternating blocks with
24s duration \cite{kaiser2010neural}. The fMRI data were processed using FSL v5.0.8 in the following pipeline: a) motion correction with MCFLIRT, 
b) interleaved slice timing correction, c) BET brain extraction, d) grand mean intensity normalization for the whole four-dimensional data set, 
e) spatial smoothing with $5mm$ FWHM, 
f) denoising with ICA-AROMA, g) nuisance regression for white matter and CSF, h) high-pass temporal filtering.   

\par
The  timing  of  the
corresponding blocks (BIO and SCRAM) was convolved with the default gamma function
(phase=0s, sd=3s, mean lag=6s) with temporal derivatives.
Participant-level t-statistics for contrast BIO>SCRAM were calculated for each voxel with first level analysis. This 3D t-statistic image is the input to the proposed approach. The input image is parcellated into 268 regions using the atlas from group-wise analysis \cite{shen2013groupwise}. 
\par
\begin{figure}[t]
    \centering
    \begin{minipage}{.32\textwidth}
        \includegraphics[width=1\linewidth]{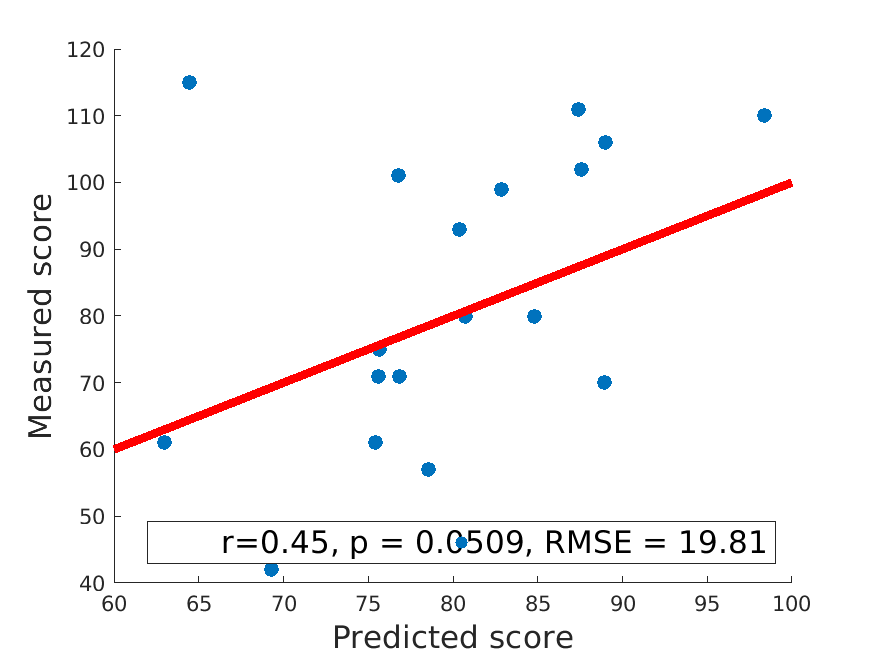}
        \caption*{
        \small{(a) Region-level elastic net model for pre-treatment SRS}}
    \end{minipage}%
    \hspace{0.015cm}
    \begin{minipage}{0.32\textwidth}
        \centering
        \includegraphics[width=1\linewidth]{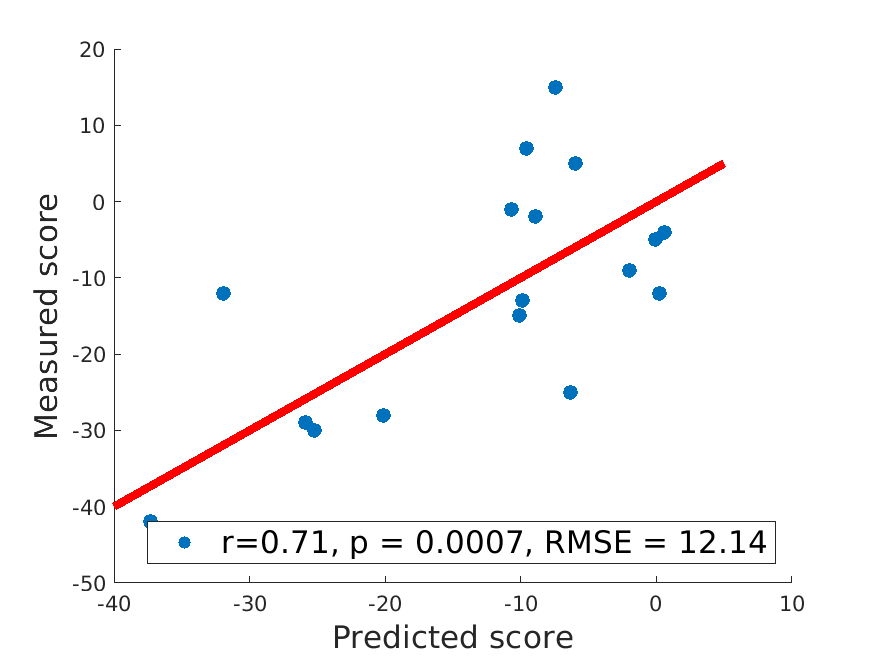}
        \caption*{\small{(b) Voxel-level random forest model for change of SRS after treatment}
        }
    \end{minipage}
    \hspace{0.015cm}
    \begin{minipage}{.32\textwidth}
        \includegraphics[width=1\linewidth]{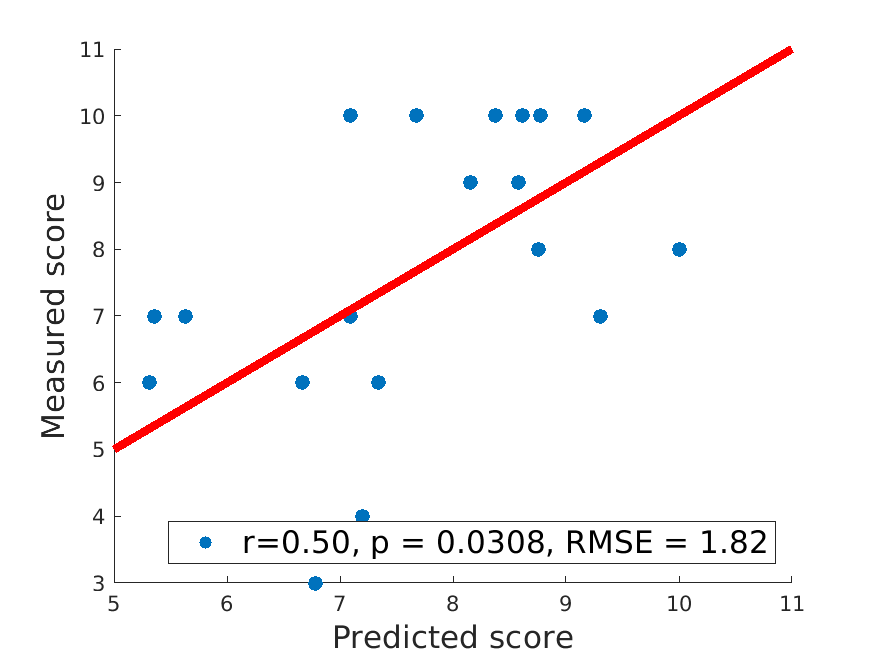}
        \caption*{\small{(c) Voxel-level random forest model for standardized ADOS}
        }
    \end{minipage}%
    \caption{
    \small{Results for various scores predicted from task fMRI, red lines are reference lines of perfect prediction $y=x$.}
    }
    \label{PRT_result1}
\end{figure}

 We tested the approach on three target scores using leave-one-out cross validation: pre-treatment SRS score, pre-treatment standardized ADOS score \cite{gotham2009standardizing}, and treatment outcome defined as the difference between pre-treatment and post-treatment SRS score. For elastic net regression, we used nested cross-validation to select parameters ($\lambda \in \{0.001, 0.01, 0.1\}$, $\alpha$ ranging from 0.1 to 0.9 with a stepsize of 0.1). Other parameters were set according to computation capability. For  random forest models, we set tree number as 2000. For region level modeling, each region was represented as the mean of 2000 bootstrapped samples from its voxels. For gradient boosted tree model after feature selection with random forest, the number of trees was set as 500. The whole process was repeated 100 times and averaged. All models were implemented in MATLAB, with default parameters except as noted above. Neurological functions of selected regions were decoded with Neurosynth \cite{yarkoni2011large}. For each experiment, results (linear correlation between predictions and measurements $r$, uncorrected p-value, root mean square error $RMSE$) of the best model are shown in Fig.  \ref{PRT_result1} and \ref{PRT_result2}.
\subsection{Resting state fMRI experiment}
We performed similar experiments on the ABIDE dataset \cite{di2014autism} using the UM and USM sites with five-fold cross validation. We selected male subjects diagnosed with ASD, resulting in 51 patients from UM and 13 patients from USM. We built models to predict the ADOS Gotham total score from voxel-mirrored homotopic connectivity images \cite{zuo2010growing}. We set parameters the same as in section \ref{task_experiment}. Results are shown in Fig. \ref{rest_fmri}.
\subsection{Result analysis}
Training and validation datasets were independent for all experiments. The proposed two-level approach accurately selected predictive features, while elastic net and RF directly applied to the whole-brain image failed to generate predictive results in all experiments (correlation between predictions and measurements < 0.1). The proposed approach generated very high predictive accuracy on various datasets and different scores, achieving better accuracy than state-of-the-art. 
\par For SRS scores, we found no predictive models in the literature. Kaiser et al. reported regions of correlation r=0.502 \cite{kaiser2010neural} in \textit{analytical} modeling, while our \textit{predictive} model achieved r=0.45 (Fig. \ref{PRT_result1}(a)).
\par For standardized ADOS score, the best result in literature achieves r=0.51 between predictions and measurements with 156 subjects based on cortical thickness \cite{moradi2017predicting}. Our model achieved r=0.50 with 19 patients (Fig. \ref{PRT_result1}(c)) based on fMRI.
\par
\begin{figure}[!htb]
    \centering
    \begin{minipage}{.33\textwidth}
        \centering
        \includegraphics[width=1\linewidth]{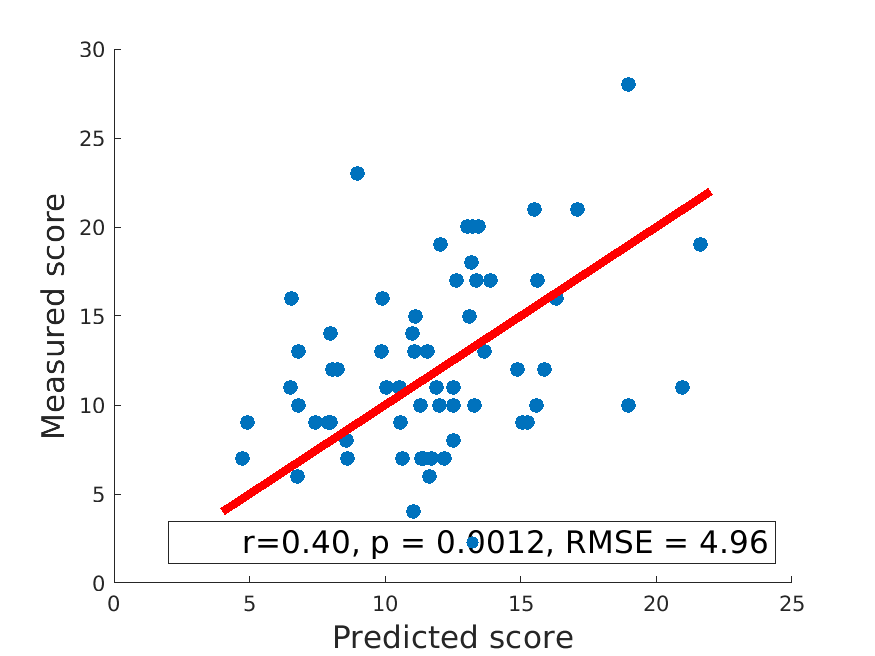}
    \end{minipage}%
    \begin{minipage}{0.43\textwidth}
        \centering
        \includegraphics[width=1\linewidth]{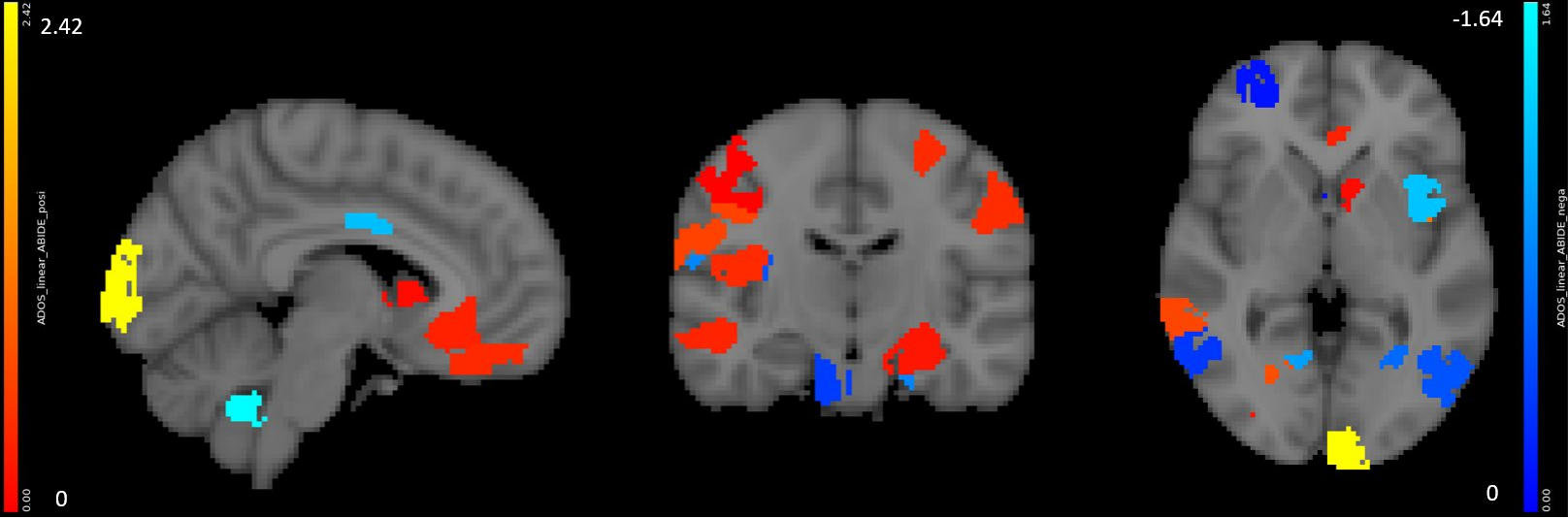}
    \end{minipage}
    \begin{minipage}{0.23\textwidth}
        \centering
        \includegraphics[width=\linewidth]{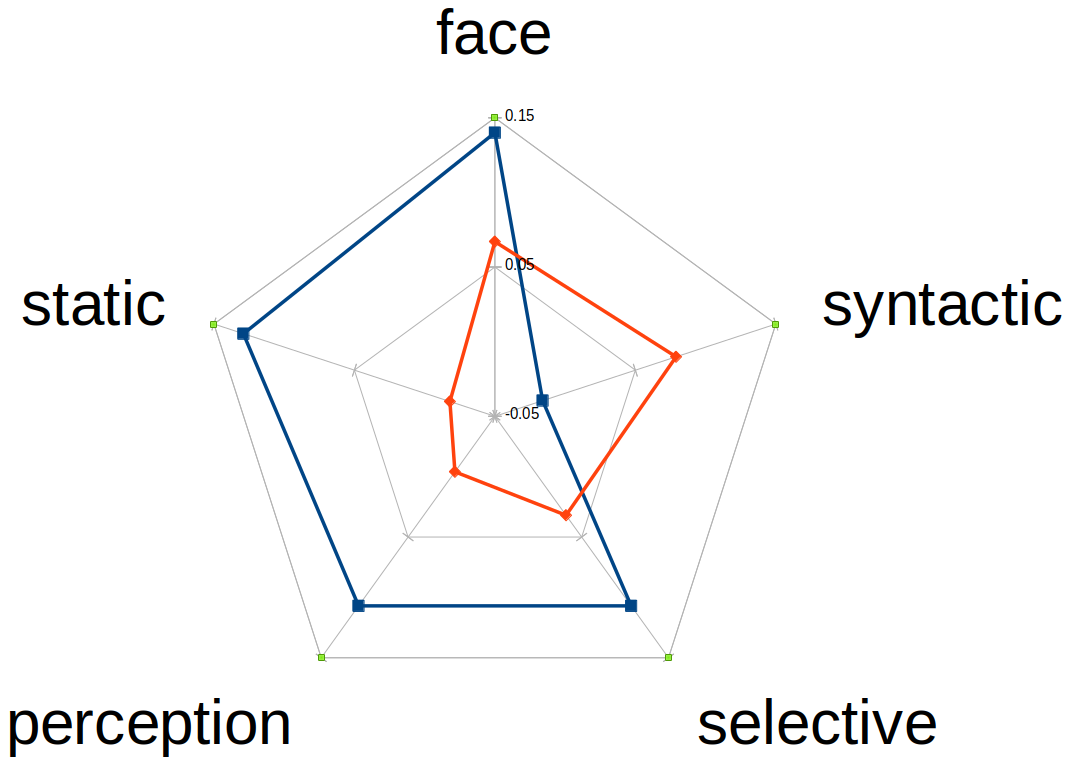}
    \end{minipage}
    \caption{
    \small{Left: Results of region-level elastic net regression model for resting-state fMRI experiment. Middle: Linear coefficients of model. Right: Functions decoded by Neurosynth, red for positive regions, blue for negative regions. }
    }
    \label{rest_fmri}
\end{figure}

 For raw ADOS score, Bj{\"o}rnsdotter et al. found no significant correlation with brain responses in fMRI scan \cite{bjornsdotter2016evaluation}. Predictive models based on structural MRI achieved correlation of r=0.362 between predictions and measurements \cite{sato2013inter}. In our experiment with resting-state fMRI, we achieved correlation r=0.40 (Fig. \ref{rest_fmri}).  
\par 
To predict treatment outcome from baseline fMRI scan, Dvornek et al. achieved correlation r=0.83 between predictions and measurements \cite{dvornekprediction}. We achieved r=0.71 (Fig. \ref{PRT_result1}(b)). However, the study by Dvornek et al. takes pre-selected regions as input and loses interpretability because it does not perform region selection. In contrast, our proposed approach takes a whole-brain image as input and can select predictive regions for interpretation and biomarker selection. Furthermore, the proposed approach is generic and any non-linear model (including Dvornek's method) can be applied.
\begin{figure}[!htb]
    \begin{minipage}{.5\textwidth}
        \centering
        \includegraphics[width=1\linewidth]{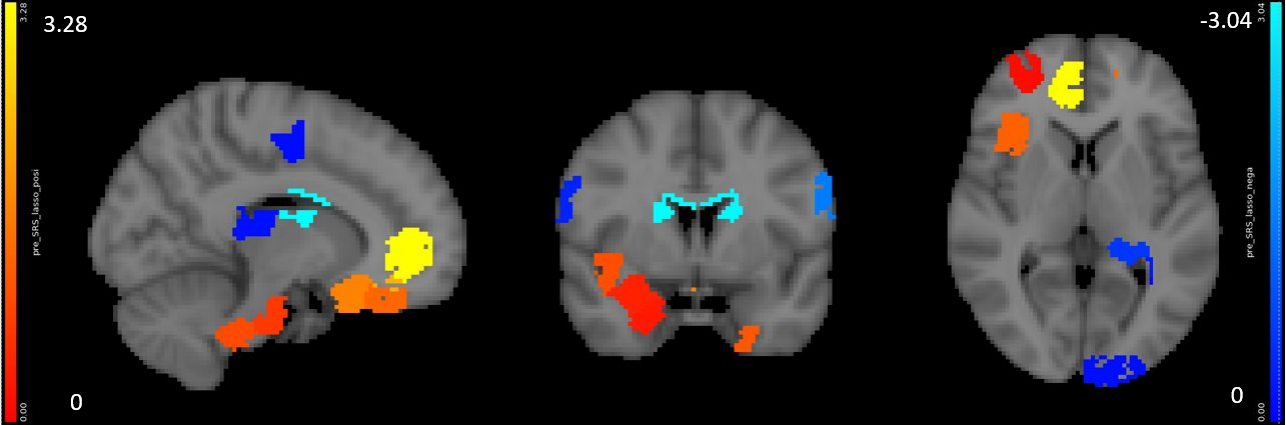}
        \caption*{\small{(a) Region-level elastic net model for pre-treatment SRS}}
    \end{minipage}%
    \hspace{0.015cm}
    \begin{minipage}{.5\textwidth}
        \centering
        \includegraphics[width=1\linewidth]{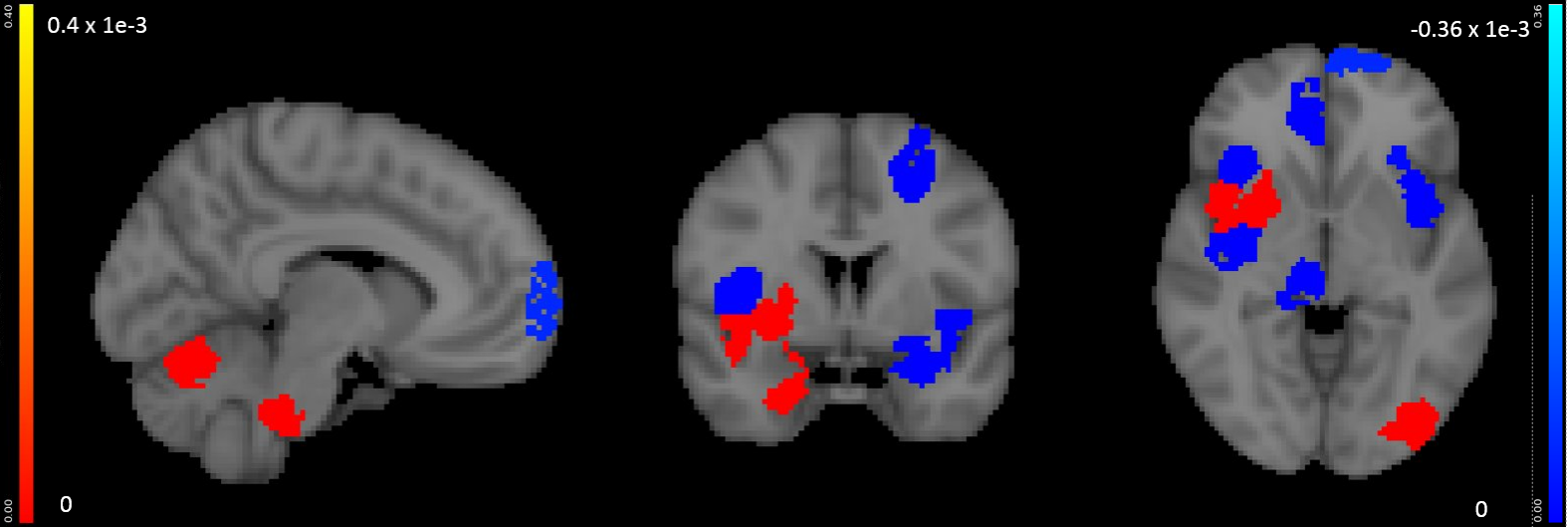}
        \caption*{ \small{(b) Voxel-level random forest model for change of SRS after treatment}
        }
    \end{minipage}%
    
    \subfloat{
    \begin{minipage}{.5\textwidth}
        \includegraphics[width=1\linewidth]{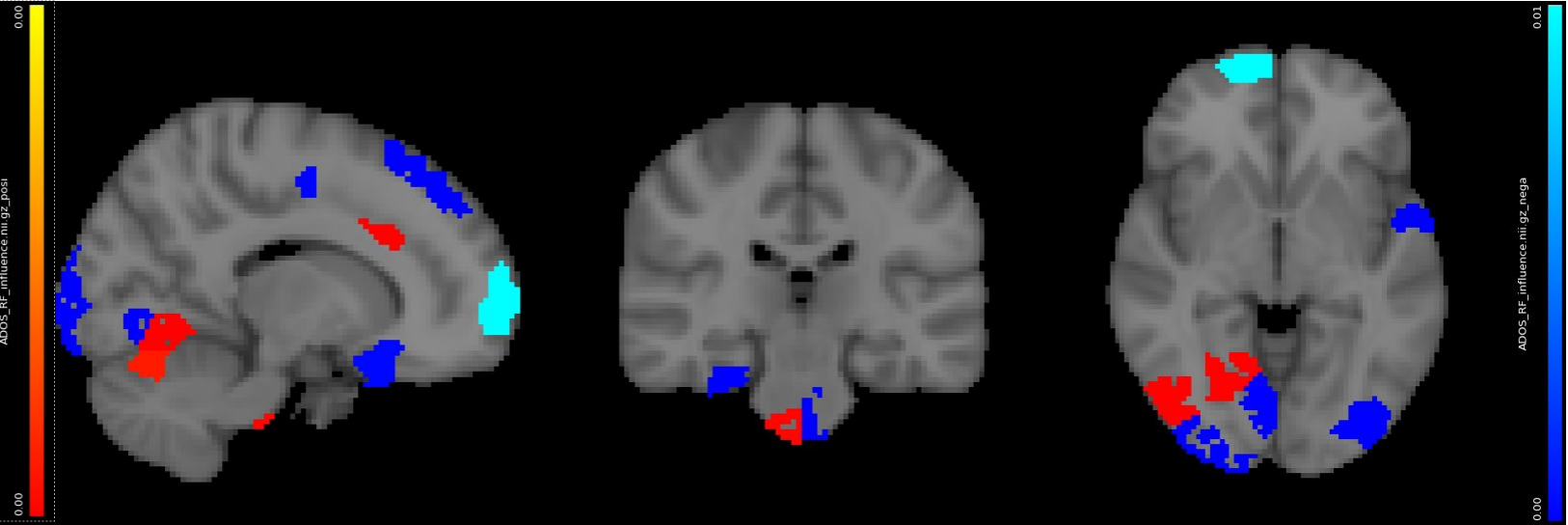}
        \caption*{\small{(c) Voxel-level random forest model for standardized ADOS}
        }
    \end{minipage}%
    }    
    \hspace{0.015cm}
    \subfloat{
    \begin{minipage}{.5\textwidth}
        \includegraphics[height=2.05cm,width=0.96\linewidth]{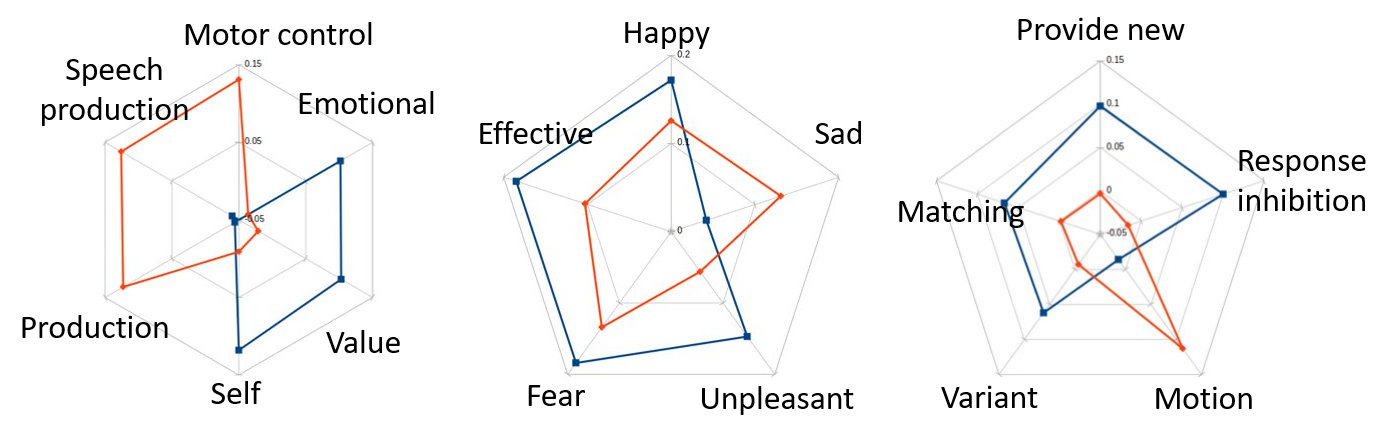}
        \caption*{\small{(d) From left to right: results of Neurosynth decoder for model in (a) (b) (c).}
        }
    \end{minipage}%
    }    
    \caption{
    \small{Regions are colored in red for positive influence and blue for negative influence. (a-c): Influence of regions for various scores based on task fMRI. (d): Functions decoded by Neurosynth.}}
    \label{PRT_result2}
\end{figure}
\par Neurosynth decoder results (Fig. \ref{PRT_result2}(d), Fig. \ref{rest_fmri} right figure) show that selected regions match the  literature \cite{kaiser2010neural}. The selected regions are slightly different across experiments due to different tasks, datasets and target measures. Many regions are shared across experiments, such as prefrontal cortex and visual cortex.
\section{Conclusion}
We propose a generic approach to build \textit{predictive} models based on fMRI images. To deal with high-dimensionality, we perform two-level variable selection: region-level modeling, and voxel-level modeling. This generic approach includes elastic net and random forest models to fit both linear and non-linear cases. The proposed approach is tested on both task-fMRI and resting-state fMRI, and validated on different scores. The proposed predictive approach achieves higher correlation than state-of-the-art predictive modeling in many experiments. Overall, the proposed approach is generic, accurate, and achieves both predictability and interpretability. 
\subsubsection*{Acknowledgement}
This research
was funded by the National Institutes of Health (NINDS-R01NS035193).
\vspace{-0.3cm}
\bibliographystyle{IEEEbib}
{\small
\bibliography{strings,refs}
}
\end{document}